# Induction of superconductivity in $Y_{0.4}Pr_{0.6}Ba_{2-x}Sr_xCu_3O_7$ system with increasing Sr substitution

V.P.S. Awana[*], M.A. Ansari, Anurag Gupta, R.B. Saxena, and H. Kishan

National Physical Laboratory K.S. Krishnan Marg, New Delhi 110012, India

Rajeev Rawat, V. Ganesan and A.V. Narlikar

Inter-University Consortium for DAE Facilities, University Campus, Khandwa Road,

Indore-452017, MP, India.

Devendra Buddhikot and S.K. Malik

Tata Institute of Fundamental Research Homi Bhabha Road, Mumbai 400005, India

Samples of $Y_{0.4}Pr_{0.6}Ba_{2-x}Sr_xCu_3O_7$ ($0.0 \leq x \leq 1.0$) have been synthesized by a solid- state reaction route. The samples with $x \leq 0.60$ crystallize in an orthorhombic structure (with orthorhombic distortion decreasing with increasing x), while the samples with $x=0.80$ and $1.0$ crystallize in a pseudo tetragonal structure. Resistance (R) measurements as a function of temperature (T) show that $x = 0.0$ sample is highly semiconducting with $R_{2K}/R_{300K}$ ratio of ~ 65. This ratio decreases to only ~23 and 5 for $x=0.20$ and $x=0.4$ samples, respectively. Further, the $x=0.60$ sample shows onset of a broad superconducting transition temperature ($T_c^{onset}$) at around 20 K without achieving the $T_c^{R=0}$ state down to 2 K. The $T_c^{R=0}$ state is observed in $x = 0.80$ and $1.00$ samples at around 5 and 14 K, respectively. Thermo-electric power (S), also exhibits $T_c^{S=0}$ state at around 11 K and 16 K, respectively for $x = 0.8$ and $1.00$ samples. Thermo electric power at room temperature is positive and decreases with increasing x, indicating enhanced number of mobile holes. These results demonstrate that substitution of Sr at Ba-site systematically induces metallicity and eventually superconductivity in $Y_{0.4}Pr_{0.6}Ba_{2-x}Sr_xCu_3O_7$ system. The results are explained on the basis of reduced Pr(4f) orbital hybridization with O(2p) in superconducting $Cu-O_2$ planes, resulting in delocalization of the mobile carriers. The role of decreased Pr/Ba intermixing disorder is also considered.

Key Words:   $Y_{0.4}Pr_{0.6}Ba_{2-x}Sr_xCu_3O_7$ system, Induction of High $T_c$ superconductivity and Pr(4f) hybridization.

Corresponding Author:  E-mail: awana@mail.nplindia.ernet.in

## 1. INTRODUCTION

The absence of superconductivity in Ce, Pr and Tb-based $REBa_2Cu_3O_7$ (RE=rare earth) compounds has attracted a lot of attraction [1,2]. The Tb and Ce based RE:123 compounds do not even form in the orthorhombic structure typical of RE:123 compounds [3]. The situation is particularly interesting in the case of Pr:123 where the crystal structure is exactly the same as that of the other 90K superconducting RE:123 compounds and yet no superconductivity is observed in this compound [1, 2]. In addition to the absence of superconductivity, the Pr ion in the Pr:123 possess an unusually high Néel temperature, $T_N$, of about 17K [1,2]

Several explanations have been proposed in the literature to explain the absence of superconductivity in Pr:123 and related compounds. These include hole filling (considering Pr to be in the tetravalent valence state [4,5]), hole localization [6] (considering Pr to be in the trivalent state), or pair breaking [7]. In fact, there are contradictory claims about the valence state of Pr in Pr:123 and related compounds. The hole-filling model is contradicted by the electron energy loss spectroscopy results where the total hole concentration in $Y_{1-x}Pr_x$:123 has been reported to be nearly the same as in other RE:123 compounds [6]. In another more viable model, the Pr(4f) electrons are assumed to hybridize with the carriers in the $Cu-O_2$ conduction planes. This strong hybridization can result in suppression of superconductivity, a high $T_N$, and a rather large electronic specific-heat coefficient [7,8,9]. There have been attempts earlier to induce superconductivity in Pr:123- type systems [10]. To the best of our knowledge, four reports exist in the literature which claim superconductivity in Pr:123 or related compounds; one on laser ablated $Pr_{0.5}Ca_{0.5}Ba_2Cu_3O_7$ thin film [11], second on bulk $Pr_{1-x}Ca_xSr_2Cu_{2.7}Mo_{0.3}O_7$ [12] and the third on $PrBa_2Cu_3O_7$ single crystals [13]. Unfortunately none of these results have been confirmed independently till now. Superconductivity is also reported with $T_c$ of up to 115 K, in high-pressure high-temperature [HPHT] synthesized Ca-doped Pr:123 system [14]. Although reported $T_c \sim 115$ K is doubtful, the 90 K superconductivity was reproducible [14].

In order, to induce metallic behavior in Pr:123-like systems, the p-type carriers in $Cu-O_2$ planes must be de-localized. To achieve this, the Pr(4f) hybridization with $Cu-O_2$ planes should be diminished. One viable approach could be the partial substitution of Ba by Sr. It is known that though this substitution in RE:123 decreases the $c$–lattice parameter, the same increases the RE to $Cu-O_2$ planes distance [15,16] which may lead to the reduction in above-mentioned hybridization.

Keeping this in mind, we have carried out such studies and report our results related to systematic induction of superconductivity in $Y_{0.4}Pr_{0.6}Ba_{2-x}Sr_xCu_3O_7$ system.

## 2. EXPERIMENTAL DETAILS

The $Y_{0.4}Pr_{0.6}Ba_{2-x}Sr_xCu_3O_7$ $(0.0 \leq x \leq 1)$ samples were synthesized by a solid-state reaction route from ingredients of $Y_2O_3$, $Pr_6O_{11}$, $SrCO_3$, $BaCO_3$, and CuO. Calcinations were carried out on the mixed powder at 900, 910, 915 and $925^?C$ each for 24 hours with intermediate grindings. The pressed circular pellets were annealed in flowing oxygen at $920^?C$ for 40 hours and subsequently cooled slowly to room temperature with an intervening annealing. at $600^?C$. for 24 hrs. X-ray diffraction (XRD) patterns were obtained at room temperature (MAC Science: MXP18VAHF[22]; $CuK_\alpha$ radiation). Resistivity Measurements were carried out by conventional four-probe method Thermoelectric power (TEP) measurements were carried out by dc differential technique over a temperature range of 5 – 300 K, using a home made set-up. Temperature gradient of ~ 1 K was maintained throughout the TEP measurements.

## 3. RESULTS

Figure 1 shows the X-ray diffraction patterns of $Y_{0.4}Pr_{0.6}Ba_{2-x}Sr_xCu_3O_7$ $(0.0 \leq x \leq 1)$ samples. The samples with $x \leq 0.60$ crystallize in the orthorhombic structure (space group, Pmmm), and the orthorhombic distortion is found to decrease with increasing x. For $x = 0.80$ and 1.0 samples, the orthorhombic distortion is nearly zero and their x-ray patterns could be indexed on the basis of a tetragonal structure (space group P4/mmm). It may be mentioned that, through normal synthesis routes (i.e. without applying High Pressure High Temperature process), $x = 1.0$ is the solubility limit of Sr at Ba site in any RE:123 system [15]. Substantial decrease in the $c$-axis lattice parameter is observed indicating successful substitution of smaller size Sr at the Ba-site in $Y_{0.4}Pr_{0.6}Ba_{2-x}Sr_xCu_3O_7$ system. Orthorhombic distortion, i.e. b-a, decreases with x, which is in agreement with previous reports on Ba-site Sr substituted RE:123 systems [15,16]. Small impurity is seen for $x = 0.80$ and 1.00 samples the line from which is marked by (*) in the X-ray patterns shown in Fig. 1. The splitting of [013] and [103] planes reverses from low angle low intensity to low angle higher intensity for $x = 0.80$ and 1.0 samples. Hence these samples were indexed in tetragonal structure (space group P4/mmm) with indexing [013] and [110] for the main peaks. Lattice parameters a, b and c are listed in Table 1 for all the samples.

Resistance versus temperature (R vs. T) plots for $Y_{0.4}Pr_{0.6}Ba_{2-x}Sr_xCu_3O_7$ ($0.0 \leq x \leq 1$) samples are shown in Fig. 2. The x = 0.0 sample is semiconducting down to 2 K, for which the $R_{2K}/R_{300K}$ ratio is nearly 65, which is in agreement with previous reports [1-7]. The x = 0.20 and 0.40 samples also exhibit semiconductor like behaviour but with somewhat reduced $R_{2K}/R_{300}$ ratios of nearly 23 and 5, respectively. This implies that the conduction improves significantly after substitution of Sr at the Ba site in this system. For x = 0.60 sample, the normal state conduction is mostly metallic in nature with a slight upturn before the superconducting onset ($T_c^{onset}$) at around 20 K. The samples with x = 0.80 and 1.0, exhibits metallic behavior in the entire temperature range between room temperature and $T_c^{onset}$. Further these two samples show zero resistance state at $T_c^{(R=0)}$ of around 5 K and 14 K, respectively. Thus one infers from Fig. 2, that both room temperature conductivity and normal state (above $T_c^{onset}$) conduction process of $Y_{0.4}Pr_{0.6}Ba_{2-x}Sr_xCu_3O_7$ system improves with increasing Sr concentration. Room temperature resistivity ($\rho_{290K}$) values of the $Y_{0.4}Pr_{0.6}Ba_{2-x}Sr_xCu_3O_7$ system for various x values are given in Table 1, and are found to systematically decrease with increase in x. Eventually, superconductivity is induced by Sr substitution at Ba site in $Y_{0.4}Pr_{0.6}Ba_{2-x}Sr_xCu_3O_7$ system.

Figure 3 shows magneto-transport behaviour of x = 0.80 superconducting sample in applied fields (H) ranging up to 0.50 Tesla. In zero external applied magnetic field, $T_c^{onset}$ is observed at around 26 K with $T_c^{(R=0)}$ at 5 K. With increasing applied field, though the $T_c^{onset}$ remains nearly constant, $T_c^{(R=0)}$ could not be observed even for small magnetic field of 0.005 Tesla. The magneto-transport measurements for x = 1.00 sample are shown in Fig.4. This sample shows $T_c^{(R=0)}$ of 14 K with $T_c^{onset}$ at around 32 K Under applied magnetic fields of up to 0.50 Tesla, the $T_c^{(R=0)}$ decreases but $T_c^{onset}$ is nearly unchanged. The situation is similar to that observed for other HTSC compounds [3].

The results of thermoelectric power (S) measurements on $Y_{0.4}Pr_{0.6}Ba_{2-x}Sr_xCu_3O_7$ ($0.0 \leq x \leq 1$) samples are shown in Fig. 5. The value of S at room temperature ($S_{290K}$) is found to be positive for all the samples, indicating them to be predominantly hole (p) type conductors. Also the value of $S_{290K}$ decreases monotonically with increasing x (Table 1) implying that the number of mobile p-type carriers increases with increasing x in $Y_{0.4}Pr_{0.6}Ba_{2-x}Sr_xCu_3O_7$ system. For strongly correlated systems, the absolute value of S is known to be inversely proportional to the number of mobile carriers [17]. Further, on lowering the temperature, S passes through a maximum ($S_{max}$) and then decreases with further decrease in temperature. The $T(S_{max})$ decreases monotonically with increasing

x. Thermoelectric power measurements below $T(S_{max})$, exhibit transition to $T_c^{S-0}$ state at around 11 K and 16 K, respectively, for x = 0.80 and 1.00 samples consistent with resistance measurements. Extended (10-50 K) S vs. T plots for x = 0.80 and 1.00 superconducting samples are shown in the inset, exhibiting the superconducting transition of the compound. Absolute S = 0 is not seen below the superconducting transition due to the contribution from the sample holder.

## 4. DISCUSSION

The results of phase formation, resistance and thermoelectric power measurements on $Y_{0.4}Pr_{0.6}Ba_{2-x}Sr_xCu_3O_7$ (0.0 ≤ x ≤ 1) samples may be summarized as follows:

[1]. Both orthorhombic distortion (b-a) and the $c$–lattice parameter decrease monotonically with increasing x.

[2]. Temperature variation of resistance exhibits insulator (x < 0.6) to metal (x >0.6) transition and the appearance of superconductivity for x = 0.8 and 1.0 samples.

[3]. Thermoelectric power data show an increase in p-type carriers with increasing x and exhibit superconductivity for x = 0.8 and 1.0 samples.

It has been observed earlier [15,16] that when bigger size Ba ion is substituted by relatively smaller size Sr ion in any RE: 123 system, oxygen vacancies are created in $CuO_{1-\delta}$ chains giving rise to reduced orthorhombicity. Also the smaller size of Sr in comparison to Ba warrants a lower $c$–lattice parameter of the system with progressive substitution. Hence the decrease of both orthorhombicity and $c$–lattice parameter of the $Y_{0.4}Pr_{0.6}Ba_{2-x}Sr_xCu_3O_7$ with increasing x is on expected lines. It has been observed from neutron diffraction studies [15,16] that though there is an overall decrease in the $c$–lattice parameter of the system, the $CuO_2$-RE-$CuO_2$ superconducting block opens up by increasing RE-O(2) distance and the $CuO_{1-\delta}$-SrO-$CuO_2$ block squeezes. The effect is shown more clearly in Fig. 6 with the help of a schematic unit cell of RE:123. The squeezing is found to be more than the opening of the blocks and hence there is an over-all decrease in the $c$–lattice parameter of the system.

Points 2 and 3 above are concerned with the fact that insulator to metal transition and further superconductivity is achieved systematically in $Y_{0.4}Pr_{0.6}Ba_{2-x}Sr_xCu_3O_7$ system with increasing x. The Pr substitution at the RE site in RE:123 compounds destroys superconductivity and brings about an insulating normal state by basically localizing the mobile hole carriers in conducting $CuO_2$ planes [1-3]. Localization of carriers happens due to induced disorder in $CuO_2$ planes [18]. The cause of such a disorder could be either Pr and Ba-sites intermixing [19] in the structure or the hybridization of extended Pr(4f) with the O(2p) orbital in $CuO_2$ planes [8,9]. The Sr substitution in an insulating $Y_{0.4}Pr_{0.6}Ba_2SrCu_3O_7$ system reduces the localization of carriers in Cu-$O_2$ planes by substantially decreasing the disorder in them. This may happen either due to decreased concentration of Ba and hence less intermixing of Pr and Ba in the system, or due to decreased Pr(4f) hybridization with O(2p) in $CuO_2$ planes.

As discussed above, on increased substitution of Sr at the Ba site in $Y_{0.4}Pr_{0.6}Ba_{2-x}Sr_xCu_3O_7$ system, the $CuO_2$-RE-$CuO_2$ superconducting block opens up by increasing RE-$O_2$ distance and the $CuO_{1-\delta}$-SrO-$CuO_2$ block squeezes with an overall decrease in the *c*–lattice parameter. An increase in RE-$O_2$ distance will result in a decrease of Pr(4f) hybridization with O(2p) orbital. Our results clearly demonstrated that the Sr substitution at Ba–site de-localizes the carriers and induces superconductivity in a systematic way in the insulating $Y_{0.4}Pr_{0.6}Ba_2Cu_3O_7$ system. Critical concentration of Pr i.e. ($x_{cr}$) depends upon the extent of Pr(4f) hybridization with O(2p). Higher concentration of Pr is needed to suppress superconductivity in a given Y/Pr:123 system if the hybridization of Pr(4f) is relatively weak with O(2p) [20]. For example, $x_{cr}$ is higher for Y/Pr:124 ($Y_{1-x}Pr_xBa_2Cu_4O_7$) than Y/Pr:123 due to weaker Pr(4f)-O(2p) hybridization in the former [20]. Interestingly, superconductivity restored in $Y_{0.4}Pr_{0.6}Ba_{2-x}Sr_xCu_3O_7$ system is only partial (14 K) and not full (90K). The reason is that while Sr substitution for Ba helps in de-localizing the carriers by decreasing Pr(4f) hybridization with $CuO_2$ planes, the same also decreases the overall number of carriers, being similar to that as for $YBa_{2-x}Sr_xCu_3O_7$ system, due to decreased oxygen content. Further, it is not possible to replace Ba completely by Sr with normal pressure heat treatments. In fact, one can synthesize compounds like $YSr_2Cu_3O_7$ with HPHT (High Pressure High Temperature) solid - state reaction route [14,15], with superconducting $T_c$ of up to 80 K. It will be interesting to see the possibility of superconductivity in HPHT synthesized $PrSr_2Cu_3O_7$ compound.

## 5. CONCLUSION

The Sr substitution at Ba site in $Y_{0.4}Pr_{0.6}Ba_{2-x}Sr_xCu_3O_7$ ($0 \leq x \leq 1$) system brings about insulator to metal transformation and superconductivity for x = 0.80 and 1.00 samples. The results may be understood on the basis of reduced Pr(4f) orbital hybridization with O(2p) in superconducting Cu-$O_2$ planes, which results in de-localization of the mobile carriers and induction of superconductivity.

**FIGURE CAPTIONS**

Figure 1. X-ray diffraction patterns of $Y_{0.4}Pr_{0.6}Ba_{2-x}Sr_xCu_3O_7$ ($0.0 \leq x \leq 1.0$) samples. Lines from a small impurity phase in x = 0.80 and 1.00 samples are marked by (*).

Figure 2. Resistance (R) vs. Temperature (T) for $Y_{0.4}Pr_{0.6}Ba_{2-x}Sr_xCu_3O_7$ ($0.0 \leq x \leq 1.0$) samples in the temperature range of 2–300 K.

Figure 3. Resistance (R) vs. Temperature (T) for $Y_{0.4}Pr_{0.6}Ba_{1.2}Sr_{0.8}Cu_3O_7$ in various applied fields, in the temperature range of 2–300 K.

Figure 4. Resistance (R) vs. Temperature (T) for $Y_{0.4}Pr_{0.6}BaSrCu_3O_7$ in various applied fields in the temperature range of 2–300 K.

Figure 5. Thermoelectric power (S) vs. Temperature (T) for $Y_{0.4}Pr_{0.6}Ba_{2-x}Sr_xCu_3O_7$ ($0.20 \leq x \leq 1$) samples in the temperature range of 5–300 K. Inset shows the extended scale S vs. T for x = 0.80 and 1.00 samples.

Figure 6. Schematic unit cell of the $REBa_{2-x}Sr_xCu_3O_7$ system.

Table 1. Lattice parameters $a$, $b$, $c$ superconducting temperature, $T_c^{(R=0)}$ and superconducting onset temperature $T_c^{onset}$ (K) for $Y_{0.4}Pr_{0.6}Ba_{2-x}Sr_xCu_3O_7$ ($0.0 \leq x \leq 1$) compounds.

| x | Space Group | a (Å) | b (Å) | c (Å) | $S_{(290K)}$ ($\mu$V/K) | $\rho_{(290K)}$ (m$\Omega$-cm) | $T_c^{(R=0)}$ (K) | $T_c^{onset}$ (K) |
|---|---|---|---|---|---|---|---|---|
| 0.00 | Pmmm | 3.869(4) | 3.901(4) | 11.7063(2) | - | 13.40 | - | |
| 0.20 | Pmmm | 3.845(3) | 3.898(6) | 11.6759(4) | 103 | 11.70 | - | - |
| 0.40 | Pmmm | 3.844(9) | 3.879(2) | 11.6356(8) | 86 | 6.13 | - | - |
| 0.60 | Pmmm | 3.860(3) | 3.868(3) | 11.5957(2) | 51 | 4.30 | - | - |
| 0.80 | P4/mmm | 3.852(5) | 3.852(5) | 11.5659(3) | 31 | 2.45 | 5 | 26 |
| 1.00 | P4/mmm | 3.849(5) | 3.849(5) | 11.5461(6) | 23 | 1.70 | 14 | 32 |

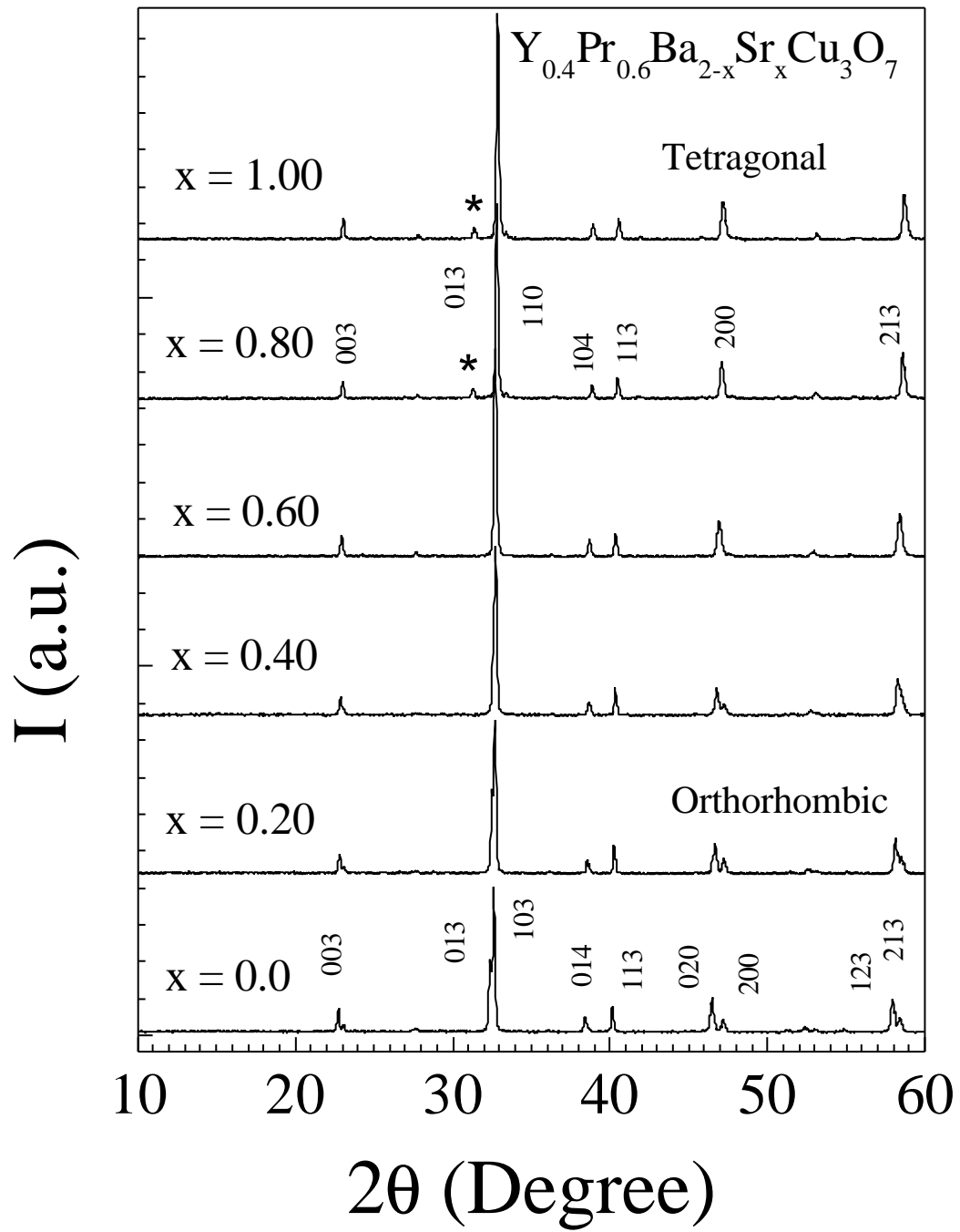

Fig.2 (Awana etal.)

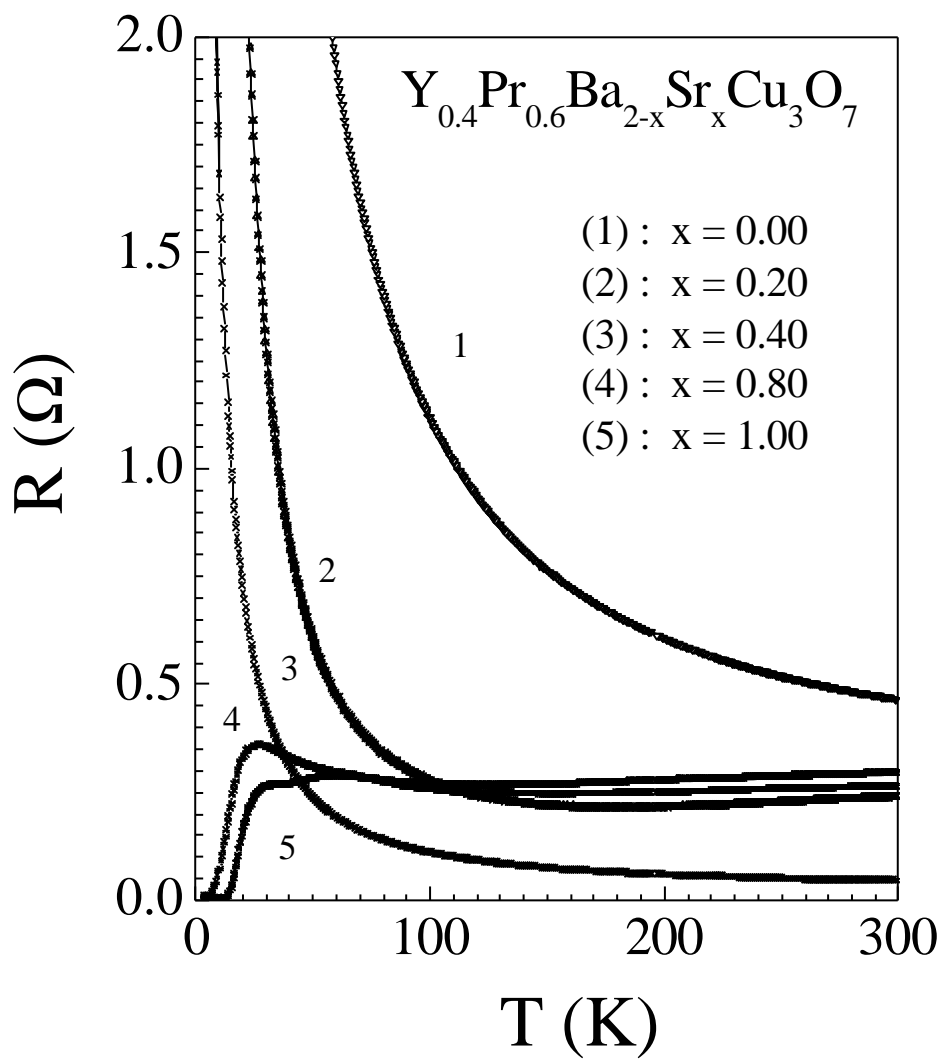

$Y_{0.4}Pr_{0.6}Ba_{2-x}Sr_xCu_3O_7$

(1) : x = 0.00
(2) : x = 0.20
(3) : x = 0.40
(4) : x = 0.80
(5) : x = 1.00

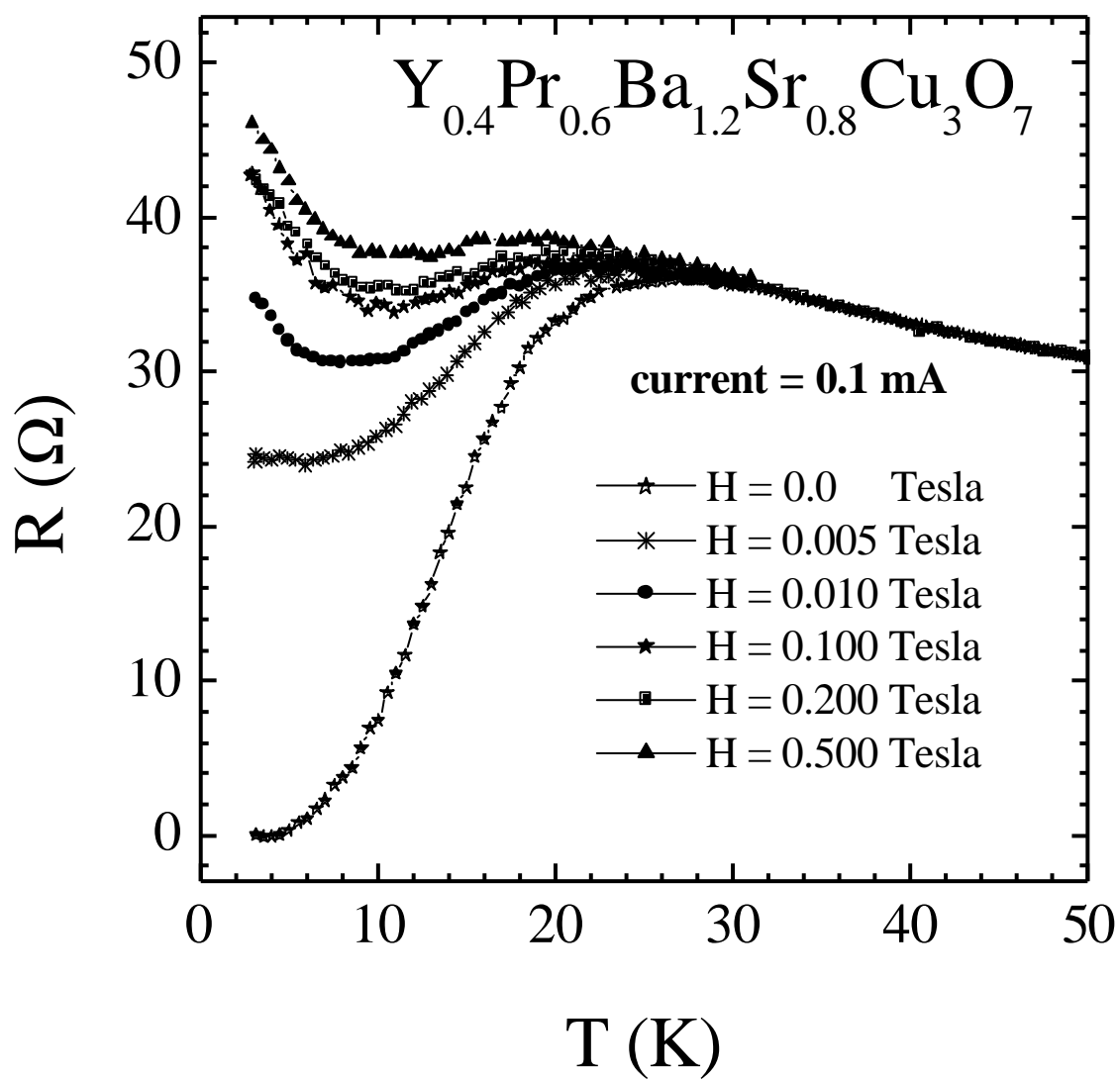

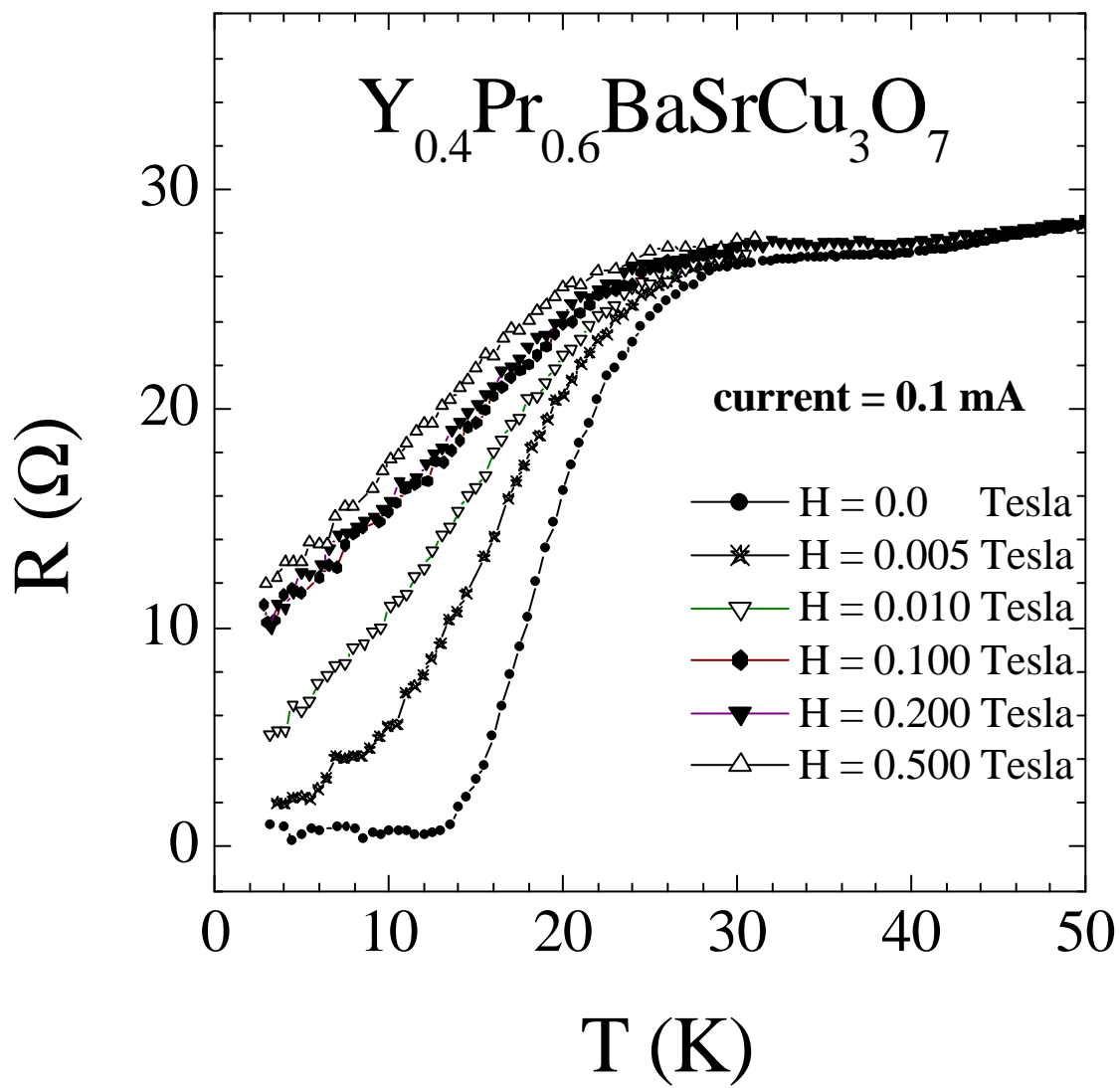

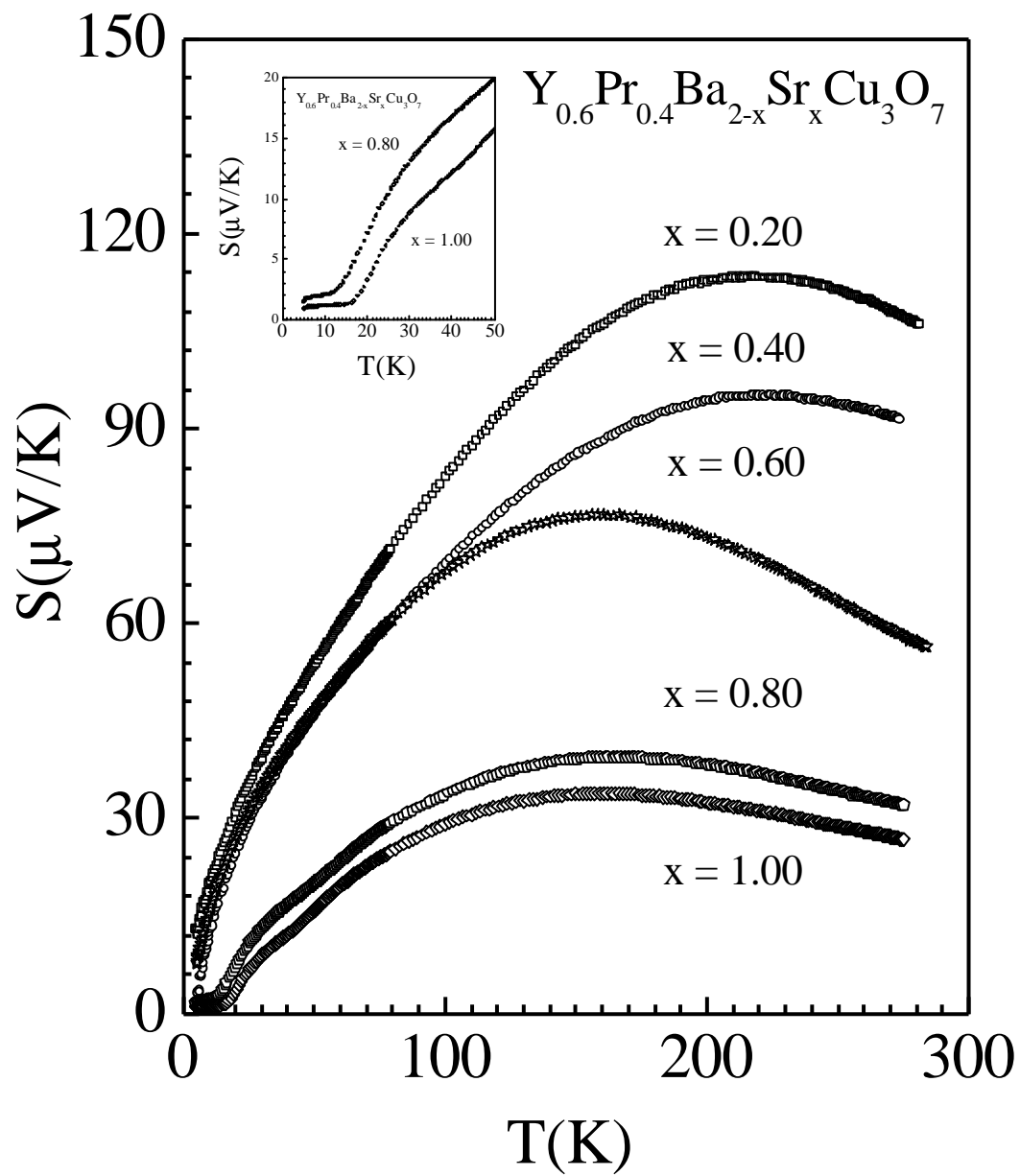

Fig.6 (Awana etal.)

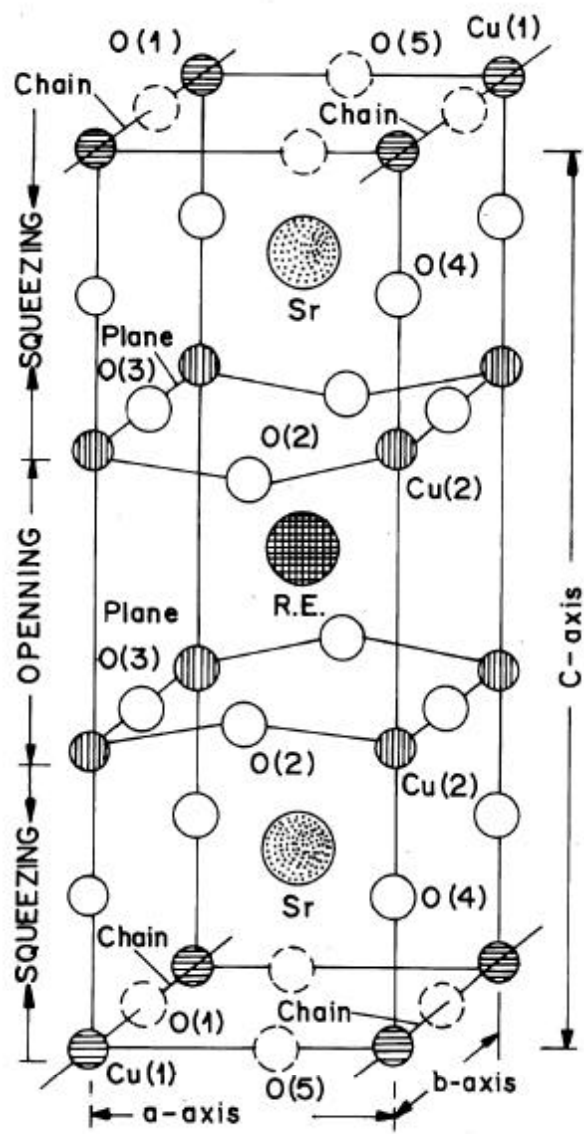